\documentclass[aps,pra,superscriptaddress,twocolumn,longbibliography]{revtex4-2}
\usepackage{graphicx}
\usepackage[T1]{fontenc}
\usepackage[colorlinks=true, allcolors=blue]{hyperref}
\usepackage{xcolor}
\usepackage{amsmath,amsfonts,amssymb}
\usepackage{booktabs}
\usepackage{multirow} 
\usepackage{array}
\usepackage{tabularx}
\usepackage{booktabs}
\begin{document}

\author{Xinyi Yuan}
\email{xinyi.yuan@aalto.fi}
\affiliation{Department of Applied Physics, Aalto University, 02150 Espoo, Finland}
\author{Grazia Salerno}
\email{grazia.salerno@unipi.it}
\affiliation{Department of Applied Physics, Aalto University, 02150 Espoo, Finland}
\affiliation{Dipartimento di Fisica "E. Fermi", Università di Pisa, Largo Bruno Pontecorvo 3, 56127 Pisa, Italy}

\title{A Group-Theoretical Framework for Local k-Space Topology and Berry Phase in 2D Photonic Systems}
\date{\today}
\begin{abstract}
Two-dimensional photonic crystals (2D PhCs) enable fine-grained control over a broad set of Bloch modes without the constraints of band occupancy and natural crystal structures, and, as intrinsically open systems, serve as versatile platforms for exploring diverse topological phenomena. Here, we develop a theoretical framework inspired by the irreducible-representation formalism in solid-state physics, while explicitly incorporating key characteristics of photonic Bloch systems, such as radiative coupling and transverse condition. Within this framework, we study the symmetry origins of local k-space topology, e.g., bound-states in the continuum and optical vortex beams, and Berry phase in two representative systems. We further analyze, from a group-theory perspective, how tailored structural designs and targeted symmetry perturbations can be exploited to manipulate these topological features. In particular, we showcase the application of the formalism to Bloch modes from distinct truncation approaches and specify the preferable regimes for each, both under a generic $n$-band configuration. The analysis can thereby be readily extended to a wide range of artificial wave crystals beyond scalar Schr\"odinger-like operators and two-level treatment.
\end{abstract}
\maketitle 
\section{Introduction}
Since the extension of topological concepts into photonics, two-dimensional photonic crystals (2D PhCs) have demonstrated great potential as platforms for exploring diverse topological phenomena \cite{haldane2008possible, raghu2008analogs, wang2008reflection, wang2009observation, lu2014topological, hsu2016bound, ozawa2019topological, price2022roadmap, tang2022topological, chan2025essay}. As a type of artificial wave crystal, 2D PhCs allow for manipulation across a broad set of Bloch modes without the constraints of band occupancy and natural crystal structures, providing enhanced flexibility to probe and engineer various Berry phase \cite{wang2016measurement, vaidya2023topological, palmer2021berry, cuerda2023observation, cuerda2024pseudospin, lehikoinen2026flat}. In addition, the nonlocality and radiative nature allow for the observation of $k$-dependent topological effects in the far-field, e.g., bound states in the continuum (BICs)~\cite{iwahashi2011higher, zhen2014topological, kodigala2017lasing, doeleman2018experimental, le2024super, yoda2020generation} and leaky modes carrying quantized orbital angular momentum~\cite{wang2020generating, zhang2023twisted, han2024generating, shen2024optical, deng2026inverse}. 

Both local $k$-space topology, such as polarization vortex, and Berry phase are closely tied to the symmetry properties of the Bloch wave functions at the high symmetry points in the Brillouin zone~\cite{xiao2010berry, po2017symmetry, blanco2020tutorial, wieder2022topological, vaidya2023topological, arjas2025topological}. As such, decomposing Bloch states at those special $k$-points into irreducible representations (IRs) could play a crucial role in building a physically transparent framework to understand the emergence of the topological phenomena in 2D PhCs. Although IR based analysis is already well established in solid state physics and has proven to be highly effective in elucidating, predicting, and classifying various band topology~\cite{slager2013space, kruthoff2017topological, bradlyn2017topological, dong2016classification, petralanda2024two}, its use in photonics is still emerging and may require adaptation to the specific properties of electromagnetic modes~\cite{doiron2022realizing, arjas2025topological}. A systematic implementation could open new avenues for the classification and design of topological features in photonics.

In particular, the level of continuum coupling, real- and $k$-space mode locality, and polarization configuration play a decisive role in determining the symmetry properties of the associated photonic systems.
They are governed by different degenerate forms of Maxwell's equations, leading to wave dynamics akin to those of different energy carriers.
The interplay between the "energy carriers", the coupling schemes and the lattice geometry actively shape the manifestation of the corresponding crystalline symmetries. 

Here, we develop a framework for adapting the IR formalism to study topological effects in passive 2D PhCs. The paper is organized as follows: we first introduce the concept of block-diagonalizing a Bloch Hamiltonian with IR in Sec.~\ref{sec:symmetryprojection}. We then discuss how in-plane modes of 2D PhCs can be engineered with the symmetry-adapted form in Sec.~\ref{sec: control}. On this foundation, we adopt different truncation approaches and apply the formalism to study $k$-dependent topological effects and Berry phase in two representative systems on the triangular lattice in Sec.~ \ref{sec: fwbic} and Sec.~\ref{sec:berryphase}, respectively. We summarize everything, followed by experimental proposals and perspectives in Sec.~\ref{sec:diss}, before concluding in Sec.~\ref{sec:ccl}.
\section{Block-diagonalizing a Hamiltonian with IRs}
\label{sec:symmetryprojection}
We use group-theoretical arguments to show how a generic Hamiltonian can be block-diagonalized and how its eigenmodes can be classified according to their irreducible representations. We consider a system described by the eigenvalue equation $H \psi = E \psi.$ 
If the system is invariant under a point group $G$, then the Hamiltonian commutes with any symmetry operation $g \in G$, where $g$ acts on the spatial coordinates, i.e., $[H, g] = 0$.
Therefore, if $\psi$ is an eigenstate of $H$ with eigenvalue $E$, then $g\psi$ is also an eigenstate with the same eigenvalue. 
For a periodic crystal, $H$ is the Bloch Hamiltonian, and the relevant symmetries can be abstracted as wallpaper groups, with the unit cells relating to crystallographic point groups. If an eigenfunction $\psi$ can be expanded in a set of linearly independent basis functions $\{\phi_i\}$ as $\psi = \sum_i c_i \phi_i$,
then the symmetry properties of $\psi$ can be characterized by how these basis functions transform under symmetry operations. In general, for a symmetry operation $g$, the basis transforms as
\begin{equation}
    g \phi_i = \sum_j D_{ji}(g)\, \phi_j,
\end{equation}
where $D(g)$ is a matrix representation of $g$ in this basis. Consequently, the action of $g$ on $\psi$ remains within the linear span of $\{\phi_i\}$, i.e., the symmetry operations act on the entire subspace spanned by the basis.
The matrices $\{D(g)\}$ thus form a reducible group representation of the eigenfunction. This representation can be further decomposed into a direct sum of irreducible representations $D^{(\alpha)}$, which provides a natural classification of the eigenstates in terms of symmetry. The decomposition is implicitly done by a similarity transformation through the projection operator 
\begin{equation}
P^{(\alpha)} = \frac{d_\alpha}{|G|} \sum_{g \in G} \chi^{(\alpha)}(g)^* D(g),
\label{eq: projection}
\end{equation}
where $|G|$ denotes total number of group elements, $d_\alpha$ and $\chi^{(\alpha)}(g)$ denote the dimension and the character of the corresponding IR $\alpha$, respectively.
By acting the projection matrix of each IR onto the original basis, one can generate the symmetry-adapted basis $ |\psi_j^{(\alpha)}\rangle = P^{(\alpha)} |\phi_i\rangle$.
Correspondingly, one can construct a unitary transformation matrix $U$ by taking the symmetry-adapted basis vectors $|\psi_j^{(\mu)}\rangle$ as its columns:
\begin{equation}
    U = \left( |\psi_1^{(1)}\rangle, \dots, |\psi_j^{(\mu)}\rangle, \dots \right).
\end{equation}
The multiplicity $n_{\alpha}$, which counts how many times an IR appears in the block-diagonalized Hamiltonian, is given by
\begin{equation}
n_\alpha = \frac{1}{|G|} \sum_{g \in G} Tr(D(g)) \cdot \chi^{(\alpha)}(g)^*.
\label{eq:reduc}
\end{equation}
Through the unitary transformation  $U^\dagger H U$, one can decompose the Bloch Hamiltonian at the $\Gamma$-point into a direct sum of independent sub-blocks as
\begin{equation}
    H_\text{block} \!=\! U^\dagger H U \!=\! \bigoplus_{\alpha} n_{\alpha} H^{(\alpha)} \!= \!
    \begin{pmatrix} 
    H^{(\Gamma_1)} & 0 & \cdots \\
    0 & H^{(\Gamma_2)} & \cdots \\
    \vdots & \vdots & \ddots 
    \end{pmatrix}.
    \label{eq:hadapted}
\end{equation}
Accordingly, the eigenfunctions, eigenvalues, and the reducible representation $D(g)$ are decomposed into the same symmetry-adapted subspaces. Notably, the same block decomposition applies directly to other high-symmetry points provided the system is effectively decoupled from the radiation continuum, as exemplified in Sec. \ref{sec:berryphase}. We demonstrate the way for analyzing the high symmetry points of open systems with non-negligible continuum coupling in Sec. \ref{sec: fwbic}.
\section{Symmetry-Adapted control of in-plane IR modes}
\label{sec: control}
\subsection{Structures and Symmetry-protected Degeneracies} 
\label{sec: structure}
\begin{figure*}
    \centering \includegraphics[width=0.85\linewidth]{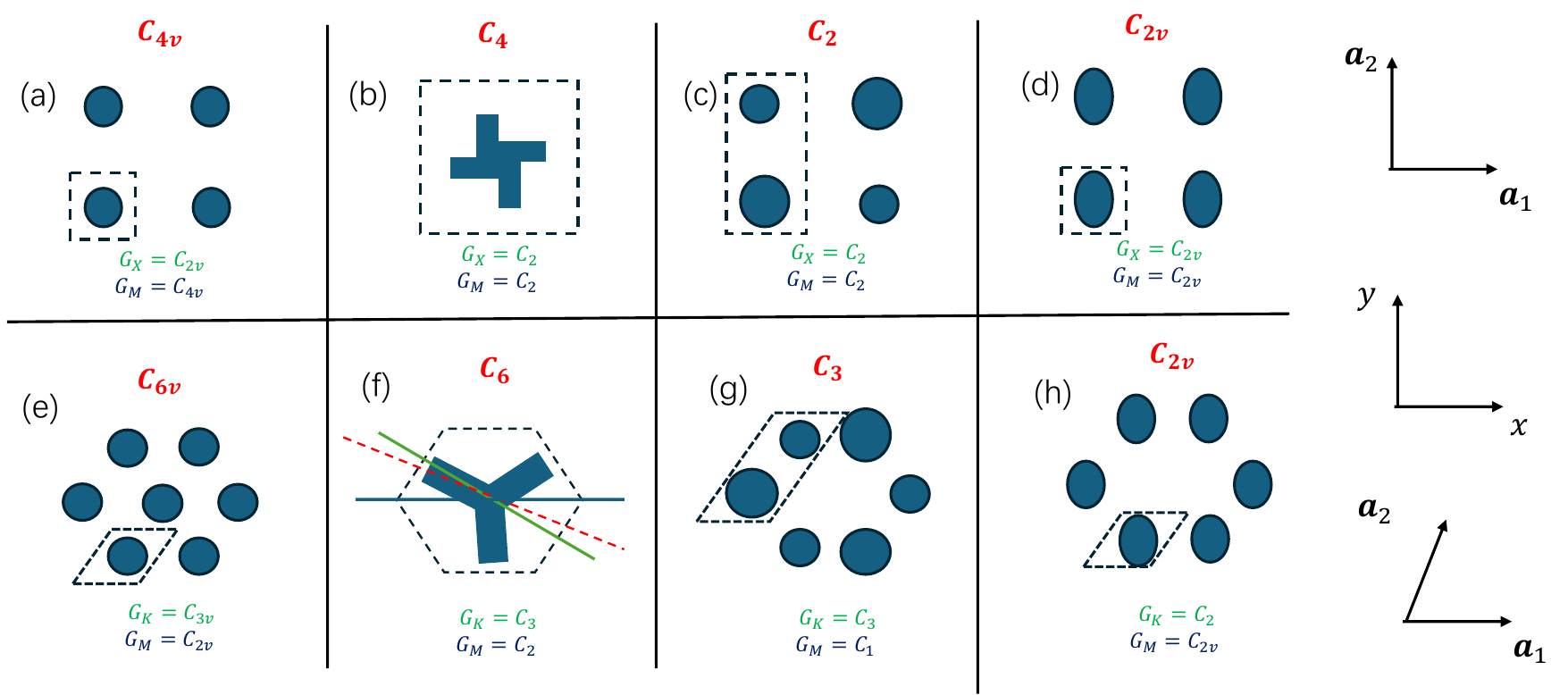}
    \caption{Representative site geometries in real space of photonic crystals, annotated with the little groups of the corresponding high-symmetry points in k-space. On the right, we show the unit vectors $\mathbf{a}_{1,2}$ of the corresponding 2D Bravais lattice. (a–d) Square lattice with different corner-site geometries: circular sites, windmill sites, circular sites of different sizes at adjacent corners, and elliptical sites, respectively. (e-h) Triangular lattice with different corner-site geometries: circular sites, $Y$-shaped sites, circular sites of different sizes at adjacent corners, and elliptical sites, respectively. Full point group class at $\Gamma$ is annotated with red color on the top, and little group classes at the corresponding high-symmetry points are annotated with green and blue colors at the bottom. The off-$\Gamma$ little groups apply only when band separability is well-defined across the whole BZ. The little group at wave vector $\mathbf{k}$ consists of all symmetry operations that satisfy $g \mathbf{k} = \mathbf{k} + \mathbf{K}$, with $\mathbf{K}$ being the reciprocal lattice vector.}
    \label{fig:littlegroups}
\end{figure*}

In the following, we discuss how representative site geometries in 2D PhCs relate to point groups defined at the $\Gamma$-point. The exemplifying structures are shown in Fig.~\ref{fig:littlegroups}. Notably, the Berry phase often originate from lifting symmetry-protected degeneracies, associated with high-dimensional IRs. In 2D systems, such high-dimensional IRs only occur in non-Abelian groups $C_{nv}$ with $n \ge 2$, e.g., Fig. \ref{fig:littlegroups}(a,e), where the reflection elements and the rotation elements of the groups don't commute. The corresponding $C_{n}$ groups, e.g., Fig. \ref{fig:littlegroups}(b,f), on the other hand, are Abelian groups possessing only 1D IRs \footnote{A group $G$ is cyclic if it is generated by a single element $g \in G$, i.e., $G = \langle g \rangle = \{ g^n \mid n \in \mathbb{Z} \}$.}. 
However, when these degeneracies are lifted by spatial symmetry perturbations, the doublets of $C_{nv}$ effectively transform into 1D complex IRs of the $C_n$ groups~\cite{doiron2022realizing}. Hence, they are, in principle, equivalent progenitors for band topology. By contrast, $C_{2v}$ is quite special in 2D due to the equivalence between $C_{2}$ operation and inversion symmetry operation
\begin{equation}
    C_2 = R(\pi) \equiv \mathcal{I}: 
\begin{pmatrix} x \\ y \end{pmatrix} 
\mapsto 
\begin{pmatrix} -x \\ -y \end{pmatrix} 
\end{equation}
where $R(\pi)$ denotes a $\pi$ rotation and $\mathcal{I}$ denotes inversion symmetry. As a result, all elements of $C_{2v}$ commute with each other, forming an Abelian group. 
Thus, all structures corresponding to $C_{2v}$ and $C_{2}$, e.g., Fig. \ref{fig:littlegroups}(c, d, h), don't satisfy the necessary condition for the emergence of Berry phase originating from symmetry-protected degeneracies. In Sec. \ref{sec:berryphase}, we extend the structure in Fig. \ref{fig:littlegroups}(a) and systematically discuss the emergence of Berry phase originating from symmetry-protected degeneracies. 

\subsection{Symmetry-constrained Mode Hybridizing Condition}
\label{sec:selectionrules}
The coupling strength between two Hermitian Bloch wave functions indexed by $i,j$ is determined by the matrix element $M_{ij} = \langle \psi_{i} | H | \psi_{j} \rangle = \int \psi_{i}^*(\tau) H(\tau) \psi_{j}(\tau) \, d\tau$. This matrix element is nonzero only if the integrand transforms as an even function under any symmetry operation. At the $\Gamma$-point, the symmetry of the integrand 
is determined by the direct product between the IR modes of the three factors. Since the Hamiltonian is inherently totally symmetric, the direct product must contain the totally symmetric representation to ensure the entire product includes $A$ or $A_1$, thereby allowing for a non-vanishing coupling between the two states. Thus, the criterion for two modes to mix is given by the direct product of the IRs of the two eigenstates as
\begin{equation}
\Gamma_{\text{mode}_1} \otimes \Gamma_{\text{mode}_2} \supset A_1.
\end{equation}
This criterion is satisfied when $A$ or $A_1$ appears in the reduction of the direct product given by Eq. \ref{eq:reduc}. In Sec. \ref{sec: fwbic}, we extend this condition to non-Hermitian regime and apply it to systematically analyze the emergence of symmetry-protected BICs and optical vortex beams.

\subsection{Band Engineering with the Symmetry Adapted Hamiltonian}
\label{sec:bandengineering}
Based on the symmetry-adapted form of the Hamiltonian in Eq.~\eqref{eq:hadapted}, the effects of different symmetry perturbations can be systematically investigated by tuning the coefficients associated with targeted IRs. Different classes of coefficients correspond to different types of symmetry-breaking mechanism. For example, real coefficients usually describe spatial symmetry perturbations, such as geometric deformations or anisotropic strain. One can apply such spatial symmetry perturbations as
\begin{equation}
    H_\text{pert} = H_\text{block} + c H^{(\Gamma_i)}
    \label{eq:tuning}
\end{equation}
where $c$ is a real number, $H^{(\Gamma_i)}$ is a matrix of the same shape as $H_{block}$ with only the $\Gamma_{i}$ elements being identities. The effect is to fine-tune the separation between the associated eigenvalue and its neighboring bands. This could be very useful when one wants to selectively permute adjacent bands in close proximity and create anti-crossings between modes that satisfy the mode-hybridizing condition. 
In Sec. \ref{sec: fwbic}, we apply this approach to study inter-block mixing under a generic $n$-band configuration. 
Moreover, the influence of applying an external field can be expressed in Eq. \ref{eq:tuning} with the non-zero elements of $cH^{(\Gamma_i)}$ being $\mathbf{F}\cdot\boldsymbol{\sigma}$, where $c$ is a real number, $\mathbf{F}$ represents the external field, and $\boldsymbol{\sigma}$ are the Pauli matrices. Specific adoption of the Pauli matrices depends on the direction of the external field. Such tuning could effectively split the two degenerate modes of a 2D IR into two complex-conjugate 1D IRs. This could be very useful when one wants to systematically lift the symmetry-protected degeneracies. In Sec. \ref{sec:berryphase}, we apply this approach to study the emergence of Berry phase under a generic $n$-band configuration.

\section{$k$-dependent Chirality Effects and Emergence of BICs}
\label{sec: fwbic}
\subsection{Representation of the In-plane Modes and Formulation of the Hamiltonian}
Symmetry-protected BICs in 2D PhCs primarily arise from the symmetry-mismatch between radiative channels and the guided modes. Based on the representativeness of the physical picture and experimental implementations, we consider a transverse electric-field (TE) configuration in a PhC slab with weak modulation and moderate degree of coupling to the continuum. We adopt a plane wave expansion (PWE) for representing the periodic part of the guided TE modes as
\begin{equation}
\mathbf{E}_{n\mathbf{k}}(\mathbf{r}) = \sum_\mathbf{G} \mathbf{c}_{\mathbf{G}} {e^{i (\mathbf{k+G})  \cdot \mathbf{r}}}.
\label{eq: fullpwe}
\end{equation}
Notably, one begins with a complete but unconstrained basis $\{ e^{i(\mathbf{k}+\mathbf{G})\cdot \mathbf{r}} \}$ in a PWE representation, spanning a full algebraic space larger than the actual physical space, i.e, $\mathcal H_{\mathrm{PWE}} \supset\mathcal H_{\mathrm{Phys}}$. Thus, the transverse condition must be explicitly imposed, effectively projecting the solutions onto the physical space as
\begin{equation}
    \mathcal{H}_{phys} = P_{\perp} \cdot \left( \mathcal{H}_{\mathrm{scalar}}  \otimes \mathbb{R}^2_{\mathrm{polarization}} \right).
    \label{eq:transverseprojection}
\end{equation}
This projection locks momentum and polarization, rendering the group representation no longer freely reducible. 

\begin{figure}[ht!]
    \centering \includegraphics[width=0.65\linewidth]{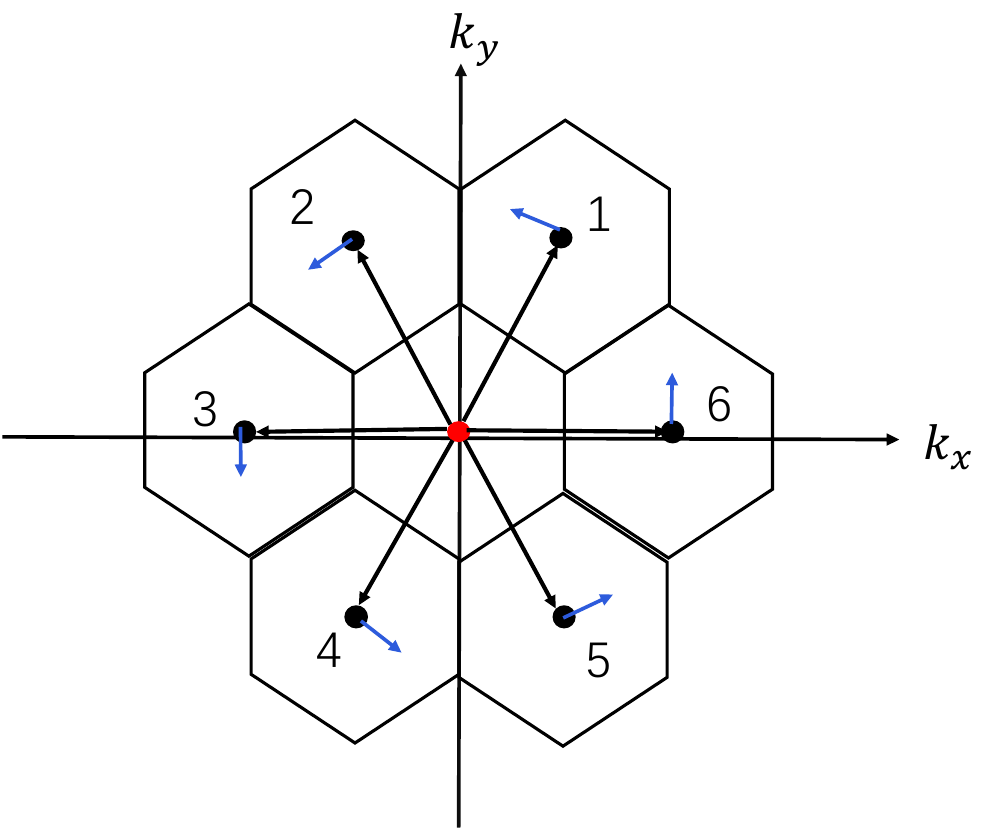}
    \caption{Schematic of the minimal constituent of the triangular model, showing the $\Gamma$-point of a triangular lattice in momentum space, the first shell of reciprocal lattice vectors as black arrows, and the polarization degrees of freedom as blue arrows.}
    \label{fig:c6vte}
\end{figure}

In the following, we showcase the procedure for determining the effective physical solutions at the $\Gamma$-point of a triangular lattice, corresponding to the $C_{6v}$ and $C_6$ groups. We consider the first shell of reciprocal lattice vectors, as schematically indicated in Fig. \ref{fig:c6vte} by the black arrow, where 
\begin{equation}
    \mathbf G_n = \frac{4\pi}{\sqrt{3}\,a_{0}}(\cos n\pi/3,\ \sin n\pi/3),
    \label{eq:gfistshell}
\end{equation}
for $n=1,\dots,6$, and $a_{0}$ is the lattice constant.

The transverse condition is applied as $\mathbf G_n \cdot \mathbf E_{\mathbf G_n} = 0$, reducing the in-plane polarization degrees of freedom to a single tangential component, schematically indicated in Fig. \ref{fig:c6vte} by the blue arrow. We define
\begin{equation}
    \mathbf t_n = \hat z \times \hat{\mathbf G}_n
= (-\sin n\pi/3,\ \cos n\pi/3),
\label{eq:tfistshell}
\end{equation}
for $n=1,\dots,6$. We approximate the electric field by truncating Eq. \ref{eq: fullpwe} with the first shell of $\mathbf{G}$ and construct the basis functions for group transformation as
\begin{equation}
    \phi_n(\mathbf r) = \mathbf t_n e^{i\mathbf G_n \cdot \mathbf r},
\end{equation}
for $n=1,\dots,6$.
We start with a $6\times6$ identity matrix $I_6$, with each column representing $\{\phi_{i} \mid i = 1, 2, \dots, 6\}$. By sequentially apply the symmetry operations $g$ of a $C_{6v}$ group, i.e., $E$, $\{C_6^k \mid k = 1, 2, \dots, 5\}$, $\sigma_{v}$, $\sigma_{d}$, one gets the permutation matrix $D(g)$ of each. With $D(g)$, one can construct the projection operators using Eq. \ref{eq: projection} and calculate the multiplicity using Eq. \ref{eq:reduc}, thereby block-diagonalizing the Hamiltonian and identifying band representations. A more detailed derivation for $C_{6v}$ group can be found in the first part of appendix~\ref{sec: appbic}, where we also highlight the difference between the deductions of $C_{6v}$ and $C_{6}$. Similarly, one can build the reducible representations $D(g)$ and determine the multiplicity of different IR modes for other commonly seen point groups under this configuration. The schematic showing the vector representation for other commonly seen point groups can be found in appendix~\ref{sec: appbic} in Fig. ~\ref{fig:firstshells}. The discussion regarding inclusion of higher diffraction orders can also be found in the second part of appendix~\ref{sec: appbic}. 

\subsection{Radiative Coupling: Emergence of BICs and Vortex Beams at the $\Gamma$-point}
The characters for reducible representations of $C_{6v}$, as discussed above, are given by $\chi_{\mathrm{shell}} = \{\text{Tr}(D(g)) \mid k = 1, 2, \dots, 6\}= (6,\ 0,\ 0,\ 0,\ -2,\ 0)$. Using the standard character table of $C_{6v}$ and the reduction formula Eq. \ref{eq:reduc}, we can deduce the first-shell in-plane modes $\Gamma_{in-plane} = A_2 \oplus B_2 \oplus E_1 \oplus E_2$. Meanwhile, the far field is an unconstrained vector field. To characterize its transformation under the point-group symmetry operations, we represent the vector nature by a $2\times2$ matrix operator, which encodes the coordinate transformation in the Euclidean plane as
\begin{equation}\begin{split}
D_{vec}(C_\beta) &= \begin{pmatrix} \cos\beta & -\sin\beta \\ \sin\beta & \cos\beta \end{pmatrix}, \\ \chi_{vec}(C_\beta) &= 2\cos\beta, \quad \chi_{vec}(\sigma) = 0.
\end{split}\end{equation}
By sequentially applying the symmetry operations $g$ of a $C_{6v}$ group, i.e., $E$, $\{C_6^k \mid k = 1, 2, \dots, 5\}$, $\sigma_{v}$, $\sigma_{d}$, one gets the permutation matrix $D_{g}$ of each. The characters are thereby given by the $\chi_{\text{far-field}} = \{\text{Tr}(D_{vec}(g) \mid k = 1, 2, \dots, 6\}= (2,\,1,\,-1,\,-2,\,0,\,0)$, coinciding with the $E_1$ row of the character table. That is to say, the far-field vector $(E_x,E_y)$ at the $\Gamma$-point transforms as $\Gamma_{\text{rad}} = E_1$. The couplings between the in-plane modes and the radiation continuum are also governed by the symmetry-based mode hybridizing condition discussed in Sec.~\ref{sec:selectionrules} as
\begin{equation}
    \Gamma_{\text{in-plane}} \otimes \Gamma_{\text{rad}} \supset A_1.
    \label{eq:selectionruleBIC}
\end{equation}
Therefore, according to the $C_{6v}$ product table, only the $E_1$ in-plane modes can couple to the far field. This indicates that $A_2, B_2, E_2$ are symmetry-protected BICs, while the $E_1$ in-plane modes are sources of optical vortex beams. This is in agreement with the IR classification in experimental observation of BICs~\cite{zhen2014topological, heilmann2022quasi, doiron2022realizing, salerno2022loss, arjas2024high}. 

This result is also in accordance with the non-Hermitian models developed in \cite{yuan2025breakdown, nguyen2025generalized}. The equivalence between the group-theoretical classification and the non-Hermitian interference model can be understood from the symmetry constraints on the radiation coupling. In the non-Hermitian description, radiation losses are incorporated through an effective Hamiltonian $H_{\mathrm{eff}}=H_0 - i W W^\dagger$, where $W$ describes the coupling between Bloch modes and radiation channels. Crucially, $W$ must respect the same symmetry, i.e., it is an intertwining operator satisfying $D_{\mathrm{rad}}(g)\, W = W\, D_{\mathrm{mode}}(g), \forall g \in G.$ As a result, $W$ belongs to the space $\mathrm{Hom}_G(\mathcal{H}_{\mathrm{mode}}, \mathcal{H}_{\mathrm{rad}})$ of symmetry-preserving maps. 

By Schur's lemma, such maps vanish unless the mode and radiation spaces match. Therefore, only modes transforming according to the same irreducible representation as the radiation continuum can couple to the far field, while all other modes satisfy $W^\dagger |\psi\rangle = 0$ and form symmetry-protected bound states in the continuum. This is precisely the same mode-hybridizing rule obtained from group theory in Eq.~\eqref{eq:selectionruleBIC},
demonstrating that the non-Hermitian interference picture is fully consistent with, and in fact a dynamical realization of, the group-theoretical constraints. That is to say, symmetry-protected BICs at the $\Gamma$ point can be equivalently interpreted as destructive interference states, since symmetry constraints enforce the vanishing of the radiation coupling matrix elements. More generally, however, interference-based BICs extend beyond symmetry protection and may arise from parameter tuning even when symmetry alone does not forbid radiation.

Similarly, one can deduce the emergence of BICs and optical vortex beams corresponding to commonly seen point groups. We conclude the far-field modes as well as the first-shell in-plane modes of representative 2D point groups, as well as the corresponding bright modes and dark modes in the first shell, as shown in appendix ~\ref{sec: appbic} in Table~\ref{tab: fulltable}. Notably, these groups can appear at all high-symmetry points. When the high-symmetry points lie above the light cone, the modes' coupling to the radiative continuum serves as a source of BICs and vortex beams. When the high-symmetry points lie below the light cone, they do not directly contribute to observable far-field phenomena. However, the analysis can still be used to predict near-field effects, e.g., polarization status, as discussed in Sec.~\ref{sec: quasibic}, and local Berry curvature concentration, as shown in Ref.~\cite{yuan2025breakdown}.

\subsection{BIC Linewidth Tuning and Spin-Duality Breaking of Vortex Beams}
\label{sec: quasibic}
With the same PWE formulation, we derive the effective non-Hermitian Hamiltonian in the vicinity the $\Gamma$-point, following~\cite{nguyen2025generalized, yuan2025breakdown}. The periodic permittivity tensor $\varepsilon(\mathbf{r})$ constituting the PhC is expanded in Fourier modes using the $\mathbf{G}$ vectors. As a result, the plane waves, i.e. guided modes of the structure, are coupled with momenta $\mathbf{k}+\mathbf{G}$.
The coupling strength between such guided modes is given by $W_{\mathbf{G},\mathbf{G}'}= \langle\mathbf{k}+\mathbf{G} | \varepsilon(\mathbf{r})|\mathbf{k}+\mathbf{G}' \rangle =
\langle u_\mathbf{G}| \varepsilon_{\mathbf{G}-\mathbf{G}'} |u_{\mathbf{G}'} \rangle$. The Fourier coefficients are symmetric $\varepsilon_{\mathbf{G}-\mathbf{G}'}= \varepsilon_{\mathbf{G}'-\mathbf{G}}$ and can be further simplified considering the $C_{6v}$ crystalline symmetry. The radiative losses arise from coupling of the Bloch components to the two transverse far-field channels, leading to an effective non-Hermitian loss matrix $ \gamma_{\mathbf{G},\mathbf{G}'}= i \langle u_\mathbf{G} | (\mathbf{t}_\mathbf{G} \cdot \mathbf{t}_{\mathbf{G}'}) \varepsilon_\mathbf{G}\varepsilon_{\mathbf{G}'} | u_{\mathbf{G}'}\rangle$. We work around $k_x \approx k_y \approx 0$ and retain only the first shell of reciprocal-lattice vectors $\mathbf{G}_n$ for $n=1,\dots,6$ in Eq.~\eqref{eq:gfistshell}. This ordering defines the basis used in the Hamiltonian below.
The total Hamiltonian $H_{\text{tot}} = H_\text{H}+ H_\text{NH}$ is composed by a Hermitian and a non-Hermitian part
\begin{align}
&H_{\text{H}} \approx 
\begin{pmatrix}
\omega_1(\mathbf{k}) & v & w & u & w^* & v \\
v & \omega_2(\mathbf{k}) & v & w & u & w^* \\
w^* & v & \omega_3(\mathbf{k}) & v & w & u \\
u & w^* & v & \omega_4(\mathbf{k}) & v & w \\
w & u & w^* & v & \omega_5(\mathbf{k}) & v \\
v & w & u & w^* & v & \omega_6(\mathbf{k})
\end{pmatrix},
\label{eq:htoth}\\
&H_{\text{NH}} \approx  i\gamma
\begin{pmatrix}
1 & \frac{1}{2} & -\frac{1}{2} & -1 & -1 & \frac{1}{2} \\
\frac{1}{2} & 1 & \frac{1}{2} & -\frac{1}{2} & -1 & -\frac{1}{2} \\
-\frac{1}{2} & \frac{1}{2} & 1 & \frac{1}{2} & -\frac{1}{2} & -1 \\
-1 & -\frac{1}{2} & \frac{1}{2} & 1 & \frac{1}{2} & -\frac{1}{2} \\
-\frac{1}{2} & -1 & -\frac{1}{2} & \frac{1}{2} & 1 & \frac{1}{2} \\
\frac{1}{2} & -\frac{1}{2} & -1 & -\frac{1}{2} & \frac{1}{2} & 1
\label{eq:htotnh}
\end{pmatrix},
\end{align}
where $\omega_n(\mathbf{k}) = \frac{\sqrt{3}}{4\pi} \mathbf{G}_n \cdot \mathbf{k}$.
The coefficients $u, v \in \mathbb{R}$ and $w \in \mathbb{R}$ for $C_{6v}$ group, while $w = w e^{i\theta} \in \mathbb{C}$ for $C_{6}$ due to mirror symmetry breaking~\footnote{Such symmetry breaking mechanism, i.e., representing mirror symmetry breaking with the complex entries, is not valid for $C_4v \rightarrow C_4$ due to the equivalence between $C_2$ operation and inversion symmetry, as discussed in Sec.~\ref{sec: structure}}. Here, $\theta$ is a dynamic angle associated with the degree of mirror symmetry breaking, e.g., the deflection angle in Fig. \ref{fig:littlegroups}(f).

The non-Hermitian part of $C_{6v}$ and $C_{6}$ groups, i.e., $H_\text{NH}$ in Eq. \ref{eq:htotnh}, is the same since the coupling terms concern only the relative acute angles between different diffraction orders. Moreover, it is invariant under both rotation and mirror operations, exhibiting $C_{6v}$ symmetric. It is strictly block-diagonalized on the symmetry-adapted basis as $U^\dagger H_\text{NH} U = 0^{(A_2)} \oplus 0^{(B_2)} \oplus 3 I_2^{(E_1)} \oplus 0^{(E_2)}$, implying that the non-Hermitian part acts only on the $E_1$ modes, i.e., making the diagonal couplings of the 2D subspace complex. The coupling terms of the Hermitian part, on the other hand, are dependent on the microscopic structures. When the mirror symmetry is broken, i.e., changing from $C_{6v}$ to $C_{6}$, the second nearest neighbor hopping, i.e., $w, w*$, gains conjugacy and becomes complex. This complex entry leads to mixing of different symmetry blocks and brings in vorticity. 

Consequently, the symmetry of $H_\text{tot}$ is essentially determined by the Hermitian part, i.e., site geometries and the corresponding coupling mechanisms. The non-Hermitian part $H_\text{NH}$, instead, is more significant in determining observables in the far field. Apart from symmetry properties, the lifetime of these observables, e.g, linewidth of BICs and vortices, can also be controlled by manipulating $H_\text{H}$, e.g., through fine-tuning of the site geometries or applying targeted symmetry perturbations.

\begin{figure*}
    \centering   \includegraphics[width=\linewidth]{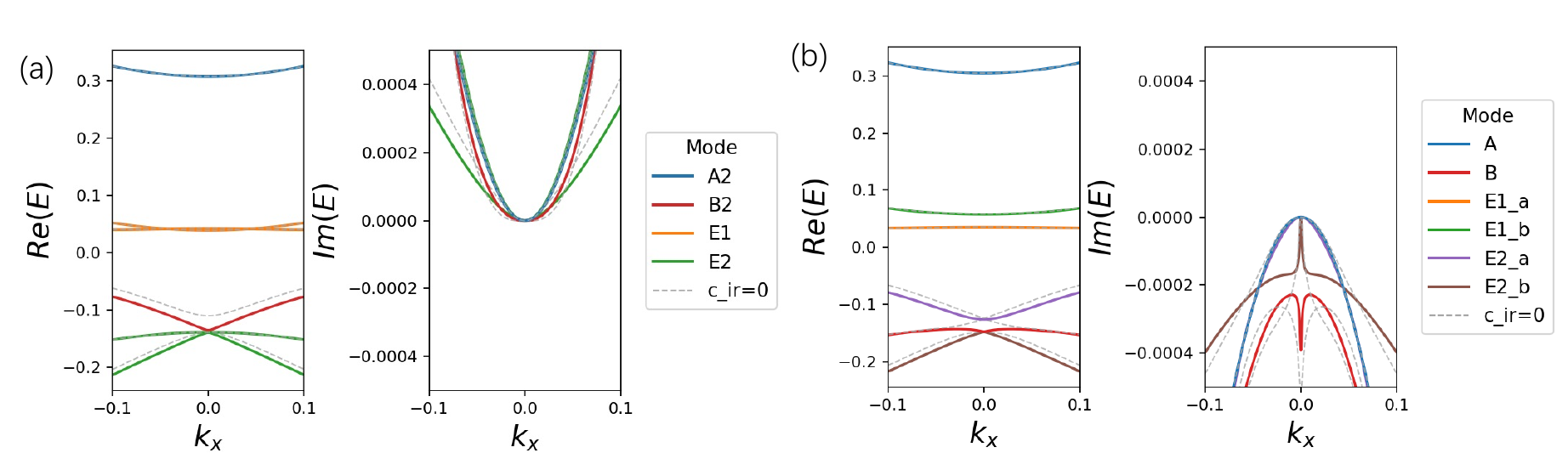}
    \caption{(a) Real and imaginary parts for the energy bands of $C_{6v}$, having used parameters $v=0.1, w= 0.05, u=0.01, \gamma=-0.005$. (b) Real and imaginary parts for the energy bands of $C_{6}$, having used parameters $v=0.1, w= 0.05\, e^{i\frac{\pi}{8}}, u=0.01, \gamma=-0.005,$. Colored solid lines and gray dashed lines represent the band structures with and without the symmetry perturbation, with the tuning coefficients being $c_{B}=c_{B_{2}}=-0.025$. The imaginary part of the leaky modes, i.e., $E_{1}$ for $C_{6v}$ and $E_{1a}$, $E_{1b}$ for $C_{6}$, are outside the plotted $y$-range.}
    \label{fig: linewidth}
\end{figure*}

By diagonalising $H_\text{tot}$, we get the band structures shown in Fig. \ref{fig: linewidth}, where the eigenmodes are classified into different IR modes. To demonstrate the effects of selective tuning of a targeted IR, we apply symmetry perturbations to the $B_2$ mode of $C_{6v}$ and the $B$ mode of $C_6$. The tuning for $C_{6v}$ modifies only the relative spectral position of the $B_2$ band, since the non-Hermitian part acts only on the $E_1$ block and no inter-block hybridization occurs. In contrast, the inter-block hybridization in $C_6$ leads to coupling among different quasi-modes. Under our parameter setting, the couplings between the $B$ mode and the two doublets are particularly pronounced. One manifestation is the hybridization between the $B$ and $E_1$ modes renders the $B$ mode a natural quasi-BIC. Another manifestation is that tuning the coefficient associated with the $B$ mode induces significant variations in the linewidth of the $B$ mode. To be more specific, we conduct purity check on the right eigenvectors, the results can be found in the second part of appendix~\ref{sec: appbic} in Table~\ref{tab:c6v_group} and Table~\ref{tab:c6_group}.

\begin{figure}[h]
    \centering \includegraphics[width=0.6\linewidth]{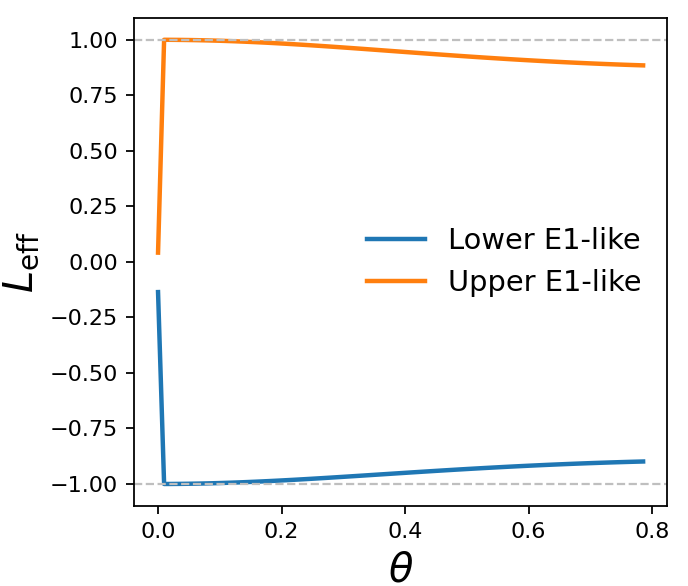}
    \caption{The influence of the structure dependent angle $\theta$ on the effective angular momentum $l_\text{eff}$ for the upper and lower quasi-$E_{1}$ modes of $C_{6}$ group. The figure is plotted with a fixed $\theta=0.05$. The degree of de-quantization of $l_\text{eff}$ increases as $\theta$ gets larger, i.e., when the mirror symmetry is progressively broken.}
    \label{fig: meff}
\end{figure}

Now we define the effective angular momentum of the right eigenvectors as $l_\text{eff}=|\langle\psi|E_{+}\rangle|^{2}-|\langle\psi|E_{-}\rangle|^{2}$, where $E_+, E_-$ are the vector representations of pure sub-modes in a 2D IR sector. This quantity measures the difference between the projections of the eigenmodes onto the two chiral bases, thereby indicating the state of circular polarization. For example, the $l_{eff}$ for the eigenstates got in Fig. \ref{fig: linewidth} are as shown in Table. \ref{tab:c6v_group} and Table. \ref{tab:c6_group}. The results show that $E_{1a}$ and $E_{1b}$ of $C_{6v}$ are by nature partially polarized, corresponding to $l_{eff}=\pm0.5$. In comparison, the $l_{eff}$ for the doulets of $C_{6}$ is close to $\pm1$, corresponding to almost fully polarized states.

The effective angular momentum $l_{\text{eff}}$ is essentially determined by a total mixing $\Theta$ composing of two parts: the dynamic coupling induced by mirror symmetry breaking, represented by $\theta$ in our parameter setting; and the geometric offset between the IR basis and the Hamiltonian's intrinsic principal axes, represented by $\phi$. A more detailed derivation on the calculation of $\Theta$ and $\phi$ can be found in the fourth part appendix~\ref{sec: appbic}. The $l_{eff}$ of $C_{6v}$ is directly given by the geometric misalignment $\phi=\frac{\pi}{6}$ since $\theta=0$. In comparison,the $l_{eff}$ of $C_6$ is essentially determined by $\theta$ as its IR basis exhibits no offset from the intrinsic eigenbasis, i.e., $\phi=0$. We show the influence of loss rate $\gamma$ and the angle $\theta$ on $l_{eff}$ of the $E_1$ modes of $C_{6}$ in Fig. \ref{fig: meff}. This indicates that, interestingly, the angular momentum carried by the experimental observables is not fully quantized; the degree of de-quantization increases as mirror symmetry is progressively broken.

\section{Inter-band Connectivity and Emergence of Berry Phase}
\label{sec:berryphase}
\subsection{Representation of the Modes and Formulation of the Hamiltonian}
Nontrivial band topology is often rooted in degeneracies at high symmetry points of Bloch crystals. These degeneracies encode the symmetry-enforced connectivity of the relevant band subspace and ultimately determine Berry-curvature distribution when the gaps are opened. Based on the representativeness of the physical picture and experimental implementations, we start with a transverse magnetic-field (TM) configuration in a passive PhC slab with strong modulation and negligible continuum coupling. We adopt a localized Wannier basis, corresponding to a tight-binding representation, to characterize the periodic part of the cavity TM modes as
\begin{equation}
\mathbf{E}_{n\mathbf{k}}(\mathbf{r}) = \sum_\mathbf{R} a_{\mathbf{R}} {e^{i \cdot  \mathbf{k}  \cdot \mathbf{(r+R)}}}.
\end{equation}
As the cavity TM modes are effectively associated with an unconstrained and localized scalar field in real space, such systems correspond directly to the standard 1D IRs of the spatial point group as 
\begin{equation}
    \Gamma = \Gamma_{spatial} \otimes A_1.
\end{equation}
Further, the little groups at other high symmetry points are automatically fixed once $\Gamma$ of the corresponding band is generated, due to the well-defined band separability across the whole BZ. 

\begin{figure}[h]
    \centering    \includegraphics[width=0.65\linewidth]{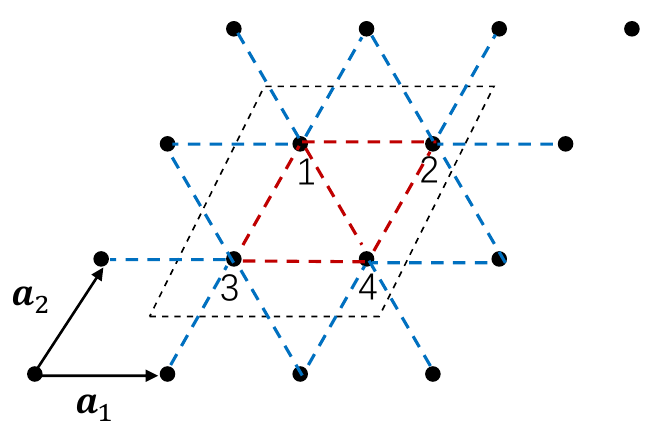}
    \caption{The model of a $2 \times 2$ triangular lattice unit cell, which can be regarded as an extension of the structure shown in Fig.\ref{fig:littlegroups}(e). The red dashed lines represent the nearest neighbor hopping within the unit cell, and the blue dashed lines represent the nearest neighbor hopping between the unit cells. The primitive vectors $\mathbf{a}_{1,2}$ are also shown.}
    \label{fig:Hc6vform2}
\end{figure}

In the following, we consider a 2D triangular lattice under a multi-band configuration, where the unit cell is composed of four particles, see Fig.~\ref{fig:Hc6vform2}.
Similar mechanisms for the formulation of the Hamiltonian in square lattices can be found in \cite{arjas2025topological}. The real-space Hamiltonian formulation reads as
\begin{equation}
    H = \sum_{m,n, \langle i,j \rangle} t \, \hat{c}^\dagger_{m,n,i} \hat{c}_{m,n,j}
\end{equation}
Here $\hat{c}_{m,n,i}^{(\dagger)}$ is the annihilation (creation) operator on the $i$-th site within the unit cell at $\mathbf{R}_{m,n}=m\mathbf{a}_1+n\mathbf{a}_2,$ with primitive vectors $\mathbf{a}_1 = \left(a_0, 0\right)$ and $\mathbf{a}_2 = \left(a_0/2, \sqrt{3}a_0/2 \right)$, where $a_0$ is the lattice constant.
We Fourier transform the lattice operators $\hat{c}_{m,n,i}=\frac{1}{\sqrt{N}}\sum_{\mathbf k}
e^{i\mathbf k\cdot \mathbf R_{m,n}} \hat{c}_{\mathbf k,i}$, such that the $k$-space Hamiltonian $H=\sum_{\mathbf k} \mathcal{H}(\mathbf k) \hat{c}^\dagger_{\mathbf k,i} \hat{c}_{\mathbf k,j}$ is
\begin{equation}
    \mathcal{H}(\mathbf{k}) = 
\begin{pmatrix}
0 & f_1(\mathbf{k}) & f_2(\mathbf{k})^* & f_{12}(\mathbf{k})^* \\
f_1(\mathbf{k})^* & 0 & g(\mathbf{k})^* & f_2(\mathbf{k})^* \\
f_2(\mathbf{k}) & g(\mathbf{k}) & 0 & f_1(\mathbf{k}) \\
f_{12}(\mathbf{k}) & f_2(\mathbf{k}) & f_1(\mathbf{k})^* & 0
\end{pmatrix}
\label{eq:TMmodel}
\end{equation}
with $f_1(\mathbf{k}) = t + t e^{-i\mathbf{k}\cdot\mathbf{a}_1}$, $f_2(\mathbf{k}) = t + t e^{-i\mathbf{k}\cdot\mathbf{a}_2}$, $f_{12}(\mathbf{k})= t + t e^{-i\mathbf{k}\cdot(\mathbf{a}_2 - \mathbf{a}_1)}$, and $g(\mathbf{k}) =  t e^{-i\mathbf{k}\cdot\mathbf{a}_1} + t e^{-i\mathbf{k}\cdot\mathbf{a}_2}$.

Notably, tight-binding uses localized orbitals centered on real-space sites. These basis functions do not individually transform according to IRs of the crystal symmetry group. Consequently, symmetry operations act nontrivially by mixing basis states, and symmetry information is not explicitly encoded in the basis. As a result, it's not as straightforward as the local treatment using the PWE basis in directly predicting which modes will appear in a given range of energy states. To identify band representations, one first formulates the real-space Hamiltonian based on corresponding real-space coupling schemes, then Fourier transforms and applies symmetry projection, i.e., Eq. \ref{eq: projection}, onto symmetry-adapted combinations after solving the eigenvalue problem. 

\subsection{Inter-Band Connectivity and Chern-Phase Progenitors}
At the $\Gamma$-point, the Hamiltonian can be block-diagonalized as $\Gamma = A_1 \oplus A_1 \oplus E_2$. The block-diagonalized form provides a natural starting point for identifying degeneracies and study topological phase transition under a specific continuous deformation. 
\begin{figure*}
    \centering \includegraphics[width=\linewidth]{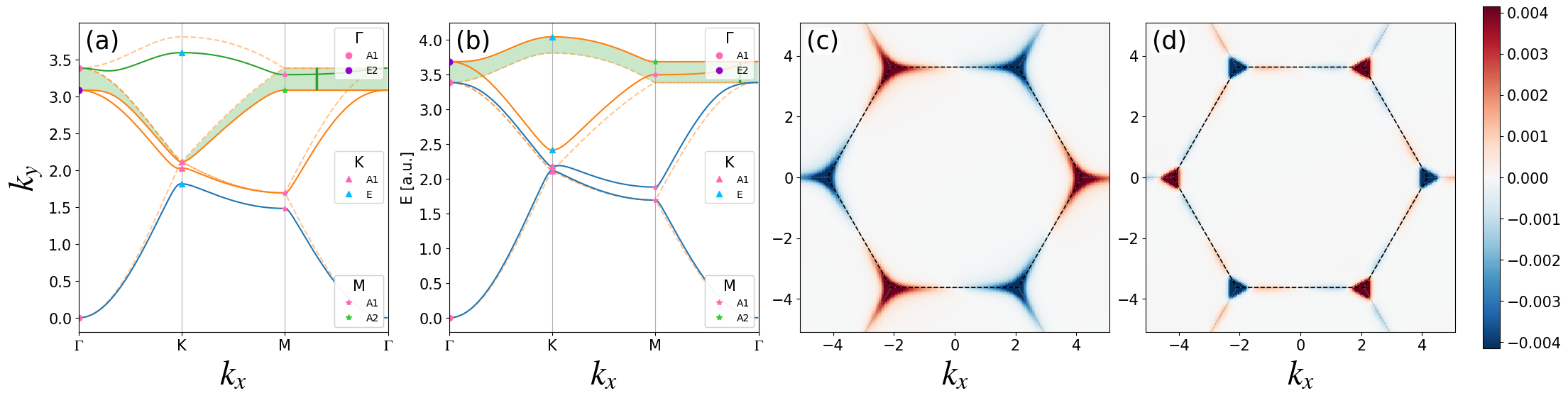}
    \caption{Energy dispersions of the illustrative system under continuous tuning of $A_{1}$ mode. The orange dashed lines indicate the empty lattice dispersion. Denote the four bands as 1–4, ordered in ascending order of energy. The subspace is partitioned into three self-consistent energetic clusters in two ways: $1,2-3,4$ and $1,2,3-4$. (a)-(b) Connectivity patterns corresponding to $c_{A_{1}}>0$ and $c_{A_{1}}<0$, respectively. Having used the parameters $c_{A_{1}}=\pm0.1$. (c)-(d) Berry curvature distribution of band 1 for the dispersions shown in (a) and (b), respectively. As the system parameters are tuned ($A_1 \to -A_1$), the band connectivity undergoes a fundamental change, resulting in a redistribution of the Berry flux in 1st BZ.}
    \label{fig: IRdispersion}
\end{figure*}
One can apply targeted symmetry perturbations on different IRs to create gaps, as discussed in Sec. \ref{sec:bandengineering}. IR perturbation modifies the inter-band connectivity of the total manifold, outlining a fertile ground for possible topological phase change. For example, by tuning the 1D IR appearing at the $\Gamma$-point, i.e., $A_1$, we identify two inter-band connectivity patterns for the illustrative system, as shown in Fig. \ref{fig: IRdispersion}(a)(b). Apart from the organization of little groups at high-symmetry points, the difference between the two types is also reflected on the Berry curvature distribution. We check the perfectly isolated band in both cases, the results are as shown Fig. \ref{fig: IRdispersion}(c)(d). We show only the results for $A_{1}$ here, since the tuning for $E_{2}$ is complementary for this specific Hamiltonian when only spatial symmetry perturbation is included.

Notably, we assign only real coefficients in this part, corresponding to a preserved TRS. In bosonic systems, e.g., classical wave crystals, TRS enforces the anti-symmetry of the Berry curvature
\begin{equation}
\Omega_n(\mathbf{k}) = - \Omega_n(-\mathbf{k}),
\end{equation}
and hence a vanishing global Chern number on the whole band
\begin{equation}
C_n = \frac{1}{2\pi} \int_{\mathrm{BZ}} \Omega_n(\mathbf{k}) \, d^2\mathbf{k} = 0.
\end{equation}
Nevertheless, net Berry curvature concentration can exist locally, e.g., at the $K, K$-point in all subplots of Fig. \ref{fig: IRdispersion}(c)(d). The distinct patterns in the Berry curvature distribution reflect a topological reconfiguration of the band subspace. The transition manifests as a realignment of the topological charge density, where the Berry flux sources and sinks are redistributed in momentum space. The inversion of the Berry curvature sign at $K,K'$-points confirms that the parameters tunes the hybridization strength between the four bands, effectively acting as a control knob for the system's topological phase. Moreover, this indicates that valley-Hall-like phases can be generalized to in multi-band Hermitian systems~\cite{wu2015scheme}. Discussion regrading inclusion of long-range couplings can be found in appendix~\ref{sec: appenberry}.

\subsection{Breaking TRS: Emergence of Chern Phases}
\begin{figure*}
    \centering \includegraphics[width=\linewidth]{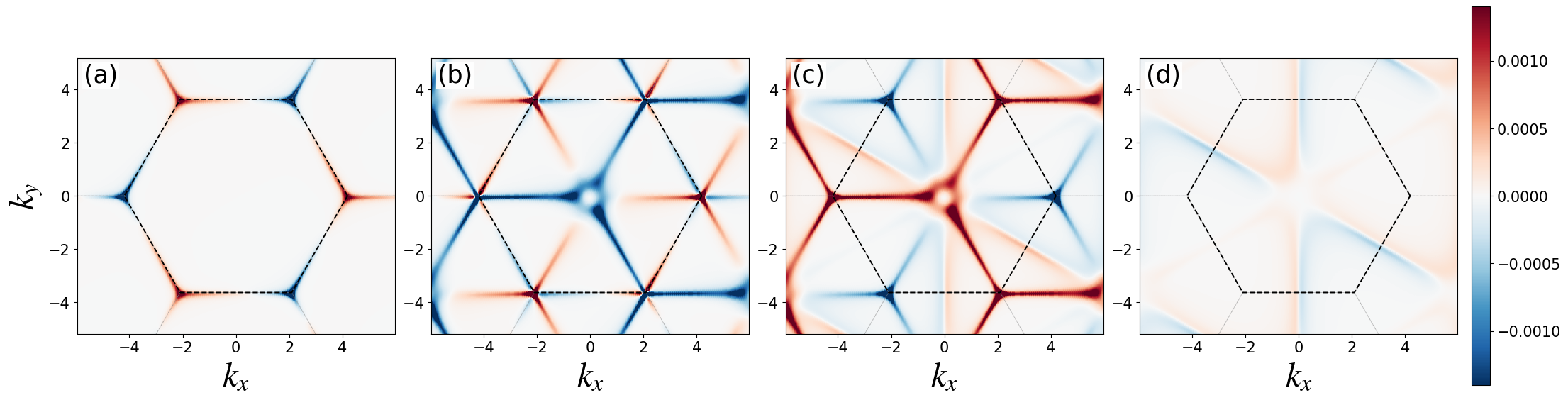}
    \caption{Results for Berry curvature distribution under representative parameter setting of applying an external magnetic field, where all the four bands are spectrally isolated. (a-d) Berry curvature distribution for band 1-4, associated with Chern numbers ${0, -1, 1, 0}$, respectively. Having used $c_{A_1}=0.1, c(\sigma_{z})= 0.05i$, where $c(\sigma_{z})$ is added to $E_2$. The isolated spectral is evolved from the connectivity pattern ${1, 2-3, 4}$, corresponding to the case where $c_{A_1}>0$ in Fig. \ref{fig: IRdispersion}.}
    \label{fig:BCTRSbroken}
\end{figure*}

To uncover Chern phases in this system, we realize TRS breaking with an external magnetic field. The mechanism is implemented through combinations of Pauli matrices, as discussed in Sec. \ref{sec:bandengineering}. In particular, we study the influence of a magnetic field applied along the $z$-axis on the 2D IR, i.e., $E_{2}$ mode appearing at the $\Gamma$-point. We examine the Berry-curvature distribution of the resulting band manifold in the fully gapped regime. The results for a representative parameter setting is as shown in Fig. \ref{fig:BCTRSbroken}.

Notably, we combine spatial perturbation and external field here. The sign and the magnitude of both $c_{A_1}$, corresponding to the structural perturbation, and $c_{\sigma_{z}}$, associated with strength and direction of the applied magnetic field, influence the distribution and accumulation of topological charges\footnote{Specific to this approach, the relative strength between $c_{A_1}$ and  $c(\sigma_{z})$ needs to be scrutinized to ensure the influence of magnetic field doesn't change the prototype after spatial perturbation.}. In particular, sign of $c_{A_1}$ controls the connectivity pattern, while sign of $c_{\sigma_{z}}$ controls chirality of the net Berry flux, i.e., sign of Chern numbers. For example, if the parameters are set to  $c_{A_1}=0.1, c(\sigma_{z})= -0.05i$, Chern numbers for the four bands are ${0, 1, -1, 0}$, respectively. Similarly, if the parameters are set to  $c_{A_1}=-0.1, c(\sigma_{z})= ± 0.05i$, Chern numbers for the four bands become ${0, 0, -1, 1}$ and ${0, 0, 1, -1}$, respectively. 

This coupling scheme represents one of the most adopted routes to band topology in photonic systems. Moreover, it can be readily generalized to related structures. Within this framework, the topological phase can be systematically constructed. The procedure starts with identifying the quasi-modes at the $\Gamma$ point and tracking their evolution under continuous spatial deformation. Such tuning modifies the inter-band connectivity, primarily through the relocation and reordering of the doublet modes, thereby generating distinct connectivity patterns of the band subspace. Once the desired band connectivity is established, breaking TRS via an external magnetic field, which introduces chirality into the symmetry-protected degeneracies, leading to the emergence of the corresponding Chern phases. 
\section{Discussion}
\label{sec:diss}
\subsection{Physical Insights: Symmetry, Topology, Nonlocality, and Non-Hermiticity} 
In Hermitian Bloch systems, PWE and Wannier-based descriptions remain formally equivalent, but they become optimal in characterizing different topological features. When applying group-theoretical analysis to the minimal models, they differ in the way of encoding symmetry and representing the underlying Hilbert space. The PWE bases encode $k$-space symmetry explicitly, since they consist of plane waves that are eigenfunctions of lattice translations and the Bloch wavevector $\mathbf{k}$ serves as a good quantum number by construction. In contrast, Wannier functions encode $k$-space symmetry implicitly, since they are based on localized orbitals centered on real-space sites, leading to nontrivial mixing of basis states when applying symmetry operations. As such, arguments given in Sec. \ref{sec: fwbic} is intrinsically more suitable for analyzing local $k$-space topology, while arguments given in Sec. \ref{sec:berryphase} is more natural for characterizing band topology. Notably, nonlocality also manifests in different ways in these two pictures. Higher order corrections of each, i.e., higher diffraction orders in PWE and long-range couplings in Wannier, embrace different details of the relevant band subspace. Nevertheless, the minimal models capture the main topological information owing to homeomorphism, as discussed in the proof given in the second part of appendix \ref{sec: appbic} and appendix \ref{sec: appenberry}.

When non-Hermiticity introduced by non-negligible continuum coupling comes into play, global band separability could become ill-defined. In general, Bloch theorem still guarantees a complete spectrum, but it does not guarantee that the bands can be globally indexed, remain single-valued, or form a smooth bundle across the whole BZ. Consequently, the little groups at the high symmetry points are not automatically fixed by the corresponding $\Gamma$-points. Nevertheless, one can still use the arguments given in Sec. \ref{sec: fwbic} to analyze the symmetry properties at high symmetry points, thus extracting the dominant topological characteristics, e.g., optical vortex beams and emergence of BICs. Besides, by building the effective local Hamiltonian, one can extract left-right eigenvectors and build far-field polarization vectors correspondingly. Moreover, one can conduct non-Hermitian AZ symmetry classification to identify gap classes and topological invariant in different regimes~\cite{kawabata_symmetry_2019}. These calculations could serve as probes for existence of BICs, exceptional points and net Berry curvature accumulation \cite{yuan2025breakdown}. Extension to higher diffractive orders can be directly understood by the minimal model, as discussed in the second part of appendix \ref{sec: appbic}. When the non-Hermiticity is introduced by external sources, however, arguments given in Sec. \ref{sec:berryphase} can also be used as a starting point for studying topological effects related to active gain-loss and time modulation, e.g., exceptional points, or Floquet Chern insulator.

\subsection{Experimental Proposals}
The arguments given in Sec. \ref{sec: fwbic} provides a pedagogical starting point for manipulating local $k$-space topology in systems with non-negligible continuum coupling. For example, by identifying the IR classification of different bands, one can apply targeted spatial symmetry perturbations to tune the BIC linewidth and engineer leakage of quasi-BIC modes. One can also control spin-duality breaking of optical vortex beams by tailoring the site geometry. Moreover, one can also manipulate the in-plane modes in proximity to observe possible topological charge redistribution in the far field.

The arguments given in Sec. \ref{sec:berryphase} provides a way to uncover Berry phase originating from symmetry-protected degeneracies in artificial supercells, when continuum coupling is negligible and band separability is well-defined across the whole BZ. Notably, this could also be a stepping stone to homeomorphic Berry phase in the subwavelength regime since the nonlocality here can be understood as natural extension of the tight-binding limit owing to homeomorphism, as discussed appendix \ref{sec: appenberry}.

Moreover, net Berry curvature concentration at the $K,K'$-points of the triangular lattice is achievable in both cases, as shown in Ref. \cite{yuan2025breakdown} and Sec. \ref{sec:berryphase}. This indicates that generalized valley-Hall-like phases are possible in lossy PhC slabs and multi-band Hermitian systems under both local and nonlocal limits. Thus, valley effects, despite its weak topological nature, could provide a pathway towards on-chip routing in a broad range of passive 2D photonic systems.

\begin{figure*}
    \centering
    \includegraphics[width=1.\linewidth]{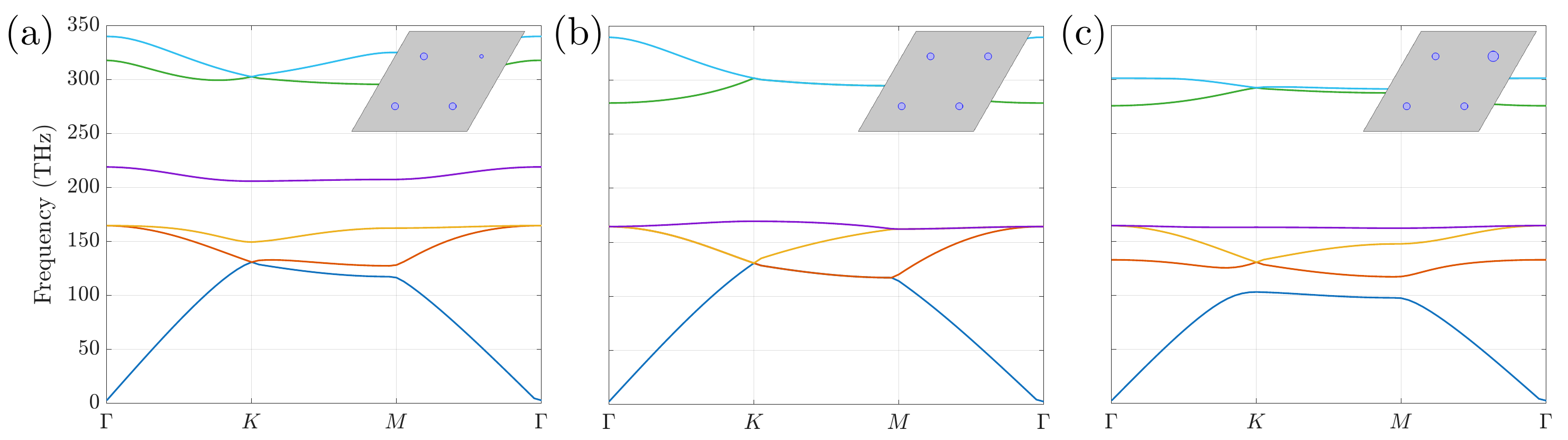}
    \caption{Full-wave COMSOL band structure of a 2D PhC with four circular dielectric sites per unit cell. The refractive index of the dielectric is $n=8$, and the lattice constant is $a_0=1\mu\mathrm{m}$.
    The rods are embedded in air and have radius $r_0=0.03~\mu\mathrm{m}$, except one whose radius is $r=r_0(1+\delta r)$, with $\delta r = -0.5$ in (a), $\delta r = 0$ in (b), $\delta r = 0.5$ in (c). Only TM out-of-plane $E_z$-polarized modes are shown. This configuration is homomorphic to the simple model in Eq.~\eqref{eq:TMmodel}, as the lowest four bands are well-isolated from the higher bands.}
    \label{fig:comsol}
\end{figure*}

In particular, we suggest a structure corresponding to the model proposed in Sec. \ref{sec:berryphase}, as shown in Fig. \ref{fig:comsol} with full-wave band structure simulation. We also summarize commonly seen experimental implementations for representative symmetry perturbations in appendix \ref{sec: appenexp}. They can be implemented accordingly both locally in $k$-space and across the whole BZ.

\section{Conclusions}
\label{sec:ccl}
 Based on the IR math formalism, we have applied a group-theoretical framework to study topological phenomena in 2D PhCs. We explicitly incorporated the key characteristics of photonic systems, i.e., bosonic nature, transversality, and substantial design flexibility. Based on the representativeness of the physical picture and experimental implementations, we applied the formalism to two archetypal settings and studied the associated topological features. In particular, we studied the origins of local $k$-space topology, i.e., BICs and optical vortex beams. We demonstrated how tailored structural designs and targeted symmetry perturbations can be exploited to understand and manipulate the emergence of these topological features. 
 
 On this foundation, we quantitatively investigated how to control the spin-duality breaking and BIC linewidth in the nonlocal and non-Hermitian regime by spatial perturbations. We also showcased how to uncover Berry phase with different symmetry perturbations in artificial supercells in the regime where band separability is well-defined. Our implementation provides a pedagogical starting point and could open new avenues for exploring topological phenomena in photonics, including more sophisticated manipulation of topological phase transitions in the far field, uncovering the Berry phase in multi-band and sub-wavelength systems, and generalized valley routing in flat optics.
 
\section{Acknowledgments}
We thank K. Arjas for fruitful discussions and for developing the initial version of the code used at the early stages of this project.
We also thank H.-S. Nguyen and P. Törmä for fruitful discussions. This work is part of the Finnish Centre of Excellence in Quantum
Materials. G.S. received support from the MUR - Italian Ministry of Research - under the Rita Levi Montalcini program.
\section{Appendix}
\subsection{BIC part appendix}
\label{sec: appbic}
\subsubsection{Extension to Commonly Seen Point Groups}
We start with a $6\times6$ identity matrix $I_6$, with each column representing $\{\phi_{i} \mid i = 1, 2, \dots, 6\}$. By sequentially apply the symmetry operations $g$ of a $C_{6v}$ group, i.e., $E$, $\{C_6^k \mid k = 1, 2, \dots, 5\}$, $\sigma_{v}$, $\sigma_{d}$, one gets the permutation matrix $D(g)$ of each. 
For rotation operation $C_6$, we have $\ \mathbf G_n \mapsto \mathbf G_{n+1},
\mathbf t_n \mapsto \mathbf t_{n+1}$, with $\mathbf G_n$ and $\mathbf t_n$ defined in Eq.~\eqref{eq:gfistshell}-\eqref{eq:tfistshell} of the main text, so that $C_6\,\phi_n = \phi_{n+1}$. For mirror operation $\sigma_v$, e.g., taking $\sigma_v$ as reflection across the $x$-axis, $\ \mathbf G_n \mapsto \mathbf G_{-n}$.
Since reflection has determinant $-1$, $\sigma_v:\ \mathbf t_n \mapsto -\mathbf t_{-n}$, thus $\sigma_v\,\phi_n = -\phi_{-n}$. Thus, the characters for
the conjugacy classes of $C_{6v}$ are $\chi(E) = 6$, $\chi(2C_6)=\chi(2C_3)=\chi(C_2)=0$, $\chi(3\sigma_v) = -2$, $\chi(3\sigma_d)=0$. 

\begin{figure*}
    \centering    \includegraphics[width=\linewidth]{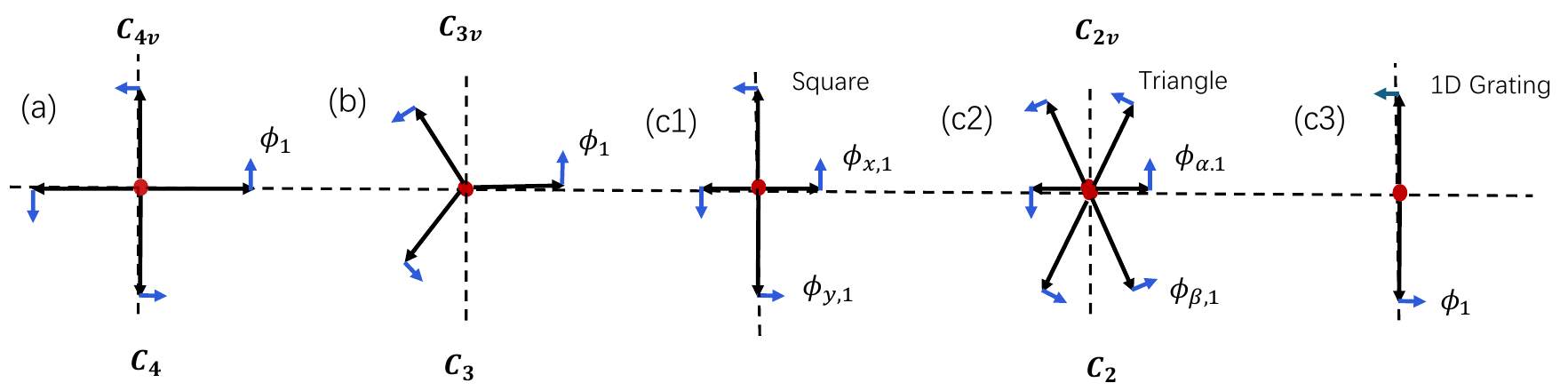}
    \caption{Schematics showing the primary plane wave space of commonly seen point groups appearing at the high symmetry points of 2D lattices, representative real-space structures can be found in Fig. \ref{fig:littlegroups}. Note every set of vector combinations $\phi_{\mathbf{G}}(\mathbf{r})=\mathbf{t}_{\mathbf{G}}\cdot e^{i\cdot \mathbf{G} \cdot \mathbf{r}}$, we have (a) Representation for the first shell of $C_{4v}$ and $C_{4}$ groups, $\mathcal H^{(1)} = \mathrm{span}\{\phi_n\}_{n=1}^4$. (b) Representation for the first shell of $C_{3v}$ and $C_{3}$ groups, $\mathcal H^{(1)} = \mathrm{span}\{\phi_n\}_{n=1}^3$. (c1) Representation for the first shell of a $C_{2v}$ and $C_{2}$ groups in square lattice, $\mathcal H^{(1)} = \mathrm{span}\{\phi_{x,n};\phi_{y,n}\}_{n=1}^2$. (c2) Representation for the first shell of $C_{2v}$ and $C_{2}$ groups in triangular lattice, $\mathcal H^{(1)} = \mathrm{span}\{\phi_{\alpha,m;\beta,n}\}_{m=1, \, n=1}^{2, \;\; \;\;\;4}$. (c3) Representation for the first shell of $C_{2v}$ and $C_{2}$ groups in 1D structures, e.g., 1D grating, $\mathcal H^{(1)} = \mathrm{span}\{\phi_{n}\}_{n=1}^2$.}
    \label{fig:firstshells}
\end{figure*}

Similarly, one can deduce the reducible representations thus the in-plane modes for other commonly seen point groups appearing at the high symmetry points. We conclude the first-shell in-plane modes of representative 2D point groups in Table. \ref{tab: fulltable}. Schematics of the vector representations are shown in Fig. \ref{fig:firstshells}. Notably, although the vector representations of $C_6$ and $C_{6v}$ are the same, they lead to different reducible representations due to difference in conjugacy classes. This can also be understood by real-space structures. Taking Fig.~\ref{fig:littlegroups}(e-f) as examples, the rotation operations cannot be grouped together for Fig.~\ref{fig:littlegroups}(f) while they can be merged for Fig.~\ref{fig:littlegroups}(e).

\begin{table*}
  \centering
\scriptsize
\renewcommand{\arraystretch}{1.1} 
\setlength{\tabcolsep}{5pt}   
\begin{tabular}{|c|c|c|c|c|}
\hline
Group & $H^{(\Gamma)}$ & Far-field & Dark & Bright
\\ \hline
$C_{6v}$ & $A_2+B_2+E_1+E_2$ & $E_1$ & $A_2+B_2+E_2$ & $E_1$ 
\\ \hline
$C_{6}$ & $\tilde{A}+\tilde{B}+\tilde{E}_{2,\{+\}}+\tilde{E}_{2,\{-\}}+\tilde{E}_{1,\{+\}}+\tilde{E}_{1,\{-\}}$ & $E_{1,\{+\}}+E_{1,\{-\}}$ & $\tilde{A}+ \tilde{B} + \tilde{E}_{2,\{+\}}+\tilde{E}_{2,\{-\}}$ & $\tilde{E}_{1,\{+\}}+\tilde{E}_{1,\{-\}}$
\\ \hline
$C_{4v}$& $A_2+B_2+E$ & $E$ & $A_2+B_2$ & $E$ 
\\ \hline
$C_{4}$ & $A+B+E $ & $E$ & $A+B$ & $E$ 
\\ \hline
$C_{2v} $ ($Sq.$) & $2A_2+B_1+B_2$ & $B_1+B_2$ & $2A_2$ & $B_1+B_2$ 
\\ \hline
$C_{2}$ ($Sq.$)& $2A+2B$ & $2B$ & $2A$ & $2B$ 
\\ \hline
$C_{3v}$ ($K$) & $A_2+E$ & $E$ & $-$ & $-$
\\ \hline
$C_{3}$ ($K$) & $\tilde{A}+\tilde{E}_{\{+\}}+\tilde{E}_{\{-\}}$ & $E_{\{+\}}+E_{\{-\}}$ & $-$ & $-$
\\ \hline
$C_{2v} $ ($1D$) & $2A_2$ & $B_1+B_{2}$ & $2A_2$ & $-$ 
\\ \hline
$C_{2} $ ($1D$) & $A+B$ & $2B$ & $A$ & $B$ 
\\ \hline
\end{tabular}
\caption{The first-shell in-plane modes and far-field IR modes of representative 2D point groups, as well as the corresponding bright modes and dark modes. The far-field modes with complex representations, i.e., $E_{\{\pm\}}$, $E_{1,\{\pm\}}$, $E_{2,\{\pm\}}$, correspond to a pair of left- and right-hand circular polarization channels. The subscript denotes the magnitude of the angular momentum. Notably, the bright modes of these channels can be sources of vortex beams. The tilted modes are intrinsically quasi modes, as discussed in Sec. \ref{sec: quasibic}. Examples of off-$\Gamma$ point and 1D group are also provided. The $C_{3v}$ and $C_{3}$ groups appearing at the $K$-point of triangular lattice are below the light cone, thus don't contribute to observables in the far-field. The illustrative structural designs corresponding to each group can be found in Fig. \ref{fig:littlegroups}. Schematics showing the primary plane wave space of commonly seen point groups appearing at the high symmetry points of 2D lattices can be found in Fig. \ref{fig:firstshells}.} 
\label{tab: fulltable}  
\end{table*}

\subsubsection{Inclusion of Higher Diffraction Orders}
The analysis of higher-energy Bloch modes can be systematically extended using the same shell-based framework, as shown above. One fixes a given high-symmetry point $k$ in the first Brillouin zone and organizes the plane-wave components according to the magnitude of $|\mathbf{k}+\mathbf{G}|$. Enlarging the representation space and allowing irreducible components absent in the lowest shell to emerge. Taking $C_{6v}$ for example, 
\begin{equation}
    H^{(n\leq 2)}=
(A_2\oplus B_2 \oplus E_1\oplus E_2)^{(1)} \oplus (A_2\oplus B_1 \oplus E_1\oplus E_2)^{(2)}.
\end{equation}
However, for each individual eigenmode, its symmetry classification remains fixed within a given IR, while higher shells only refine its structure within the same symmetry sector, i.e., 
\begin{equation}
    H = \begin{pmatrix} 
    H^{\text{shell}} & \epsilon \\ 
    \epsilon & H_{\text{rest}} 
    \end{pmatrix}
    \approx \begin{pmatrix} 
    \bigoplus H^{(1)+\dots+(n)}_{\alpha} & 0 \\ 
    0 & H_{\text{rest}}
    \end{pmatrix}.
\end{equation}
By considering more and more shells of $\mathbf G$-vectors, such shell-based symmetry classification breaks down at higher frequency approaching the continuum limit. The energy separation between different $|\mathbf{k}+\mathbf{G}|$ shells diminishes and inter-shell coupling becomes comparable to the shell spacing, leading to a mixing of multiple shell subspaces within the same energy range.

\subsubsection{Purity Check of Right Eigenvectors}
The right eigenvectors of the total Hamiltonian can be expressed in the symmetry adapted basis as $|\psi_{i}\rangle=\sum_{\alpha, i}c_{\alpha,i}|\psi_{\alpha}\rangle$. We use $|c_{\alpha,i}|^2$ as a measure to assess the IR purity: $|\psi_{i}\rangle$ is identified as a pure $\alpha$-mode when $|c_{\alpha,i}|^2=1$. The results for the purity check of the eigenstates got in Fig. \ref{fig: linewidth} are as shown in Table~\ref{tab:c6v_group} and Table~\ref{tab:c6_group}, respectively. Since the complex term introduced by the non-Hermitian part acts only on the diagonal elements of the $E_1$ block, the total Hamiltonian of $C_{6v}$ remain fully block-diagonalizable. As a result, all the eigenstates are pure states. 

On the contrary, the complex entry provided by mirror symmetry breaking hybridizes different IR blocks, leading to subspace mixing for the total Hamiltonian of $C_{6}$. Therefore, the resulting eigenstates no longer correspond to pure IR modes. Nevertheless, when the hybridization remains moderate, each eigenstate retains its own dominant IR component. In this sense, these states can be viewed as quasi-modes continuously connected to the original IR blocks. 
\begin{table*}
  \centering
  \scriptsize
  \renewcommand{\arraystretch}{1.1} 
  \setlength{\tabcolsep}{5pt}   
  \begin{tabular}{|c|c|c|c|c|c|c|c|c|}
  \hline
Band & $Re(E) (\times10^{-1})$ & $Im(E) (\times10^{-3})$ & $|c_{A_2}|^{2}$ & $|c_{B_2}|^{2}$ & $|c_{E_1}|^{2}$ & $|c_{E_2}|^{2}$ & $l_\text{eff}(E_{1})$ & $l_\text{eff}(E_{2})$ \\ \hline
1 & -1.40 & -0.0  & 0.00 & 0.00 & 0.00 & 1.00 & -0.00 & -0.50 \\ \hline
2 & -1.39 & +0.0  & 0.00 & 0.00 & 0.00 & 1.00 & -0.00 & +0.50 \\ \hline
3 & -1.10 & +0.0  & 0.00 & 1.00 & 0.00 & 0.00 & -0.00 & +0.00 \\ \hline
4 & +0.39 & +0.15 & 0.00 & 0.00 & 1.00 & 0.00 & +0.50 & +0.00 \\ \hline
5 & +0.42 & +0.15 & 0.00 & 0.00 & 1.00 & 0.00 & -0.50 & +0.00 \\ \hline
6 & +3.08 & +0.0  & 1.00 & 0.00 & 0.00 & 0.00 & +0.00 & -0.00 \\ \hline
\end{tabular}
\caption{Results for fidelity check and effective angular momentum analysis for $C_{6v}$. Having used the same parameters as in Fig. \ref{fig: linewidth} with $c_{ir}=0$. Band indices are assigned in ascending order of the real parts.}
\label{tab:c6v_group}
\end{table*}

\begin{table*}
  \centering
  \scriptsize
  \renewcommand{\arraystretch}{1.1} 
  \setlength{\tabcolsep}{5pt} 
  \begin{tabular}{|c|c|c|c|c|c|c|c|c|c|c|}
  \hline
Band & $Re(E) (\times10^{-1})$ & $Im(E) (\times10^{-3})$ & $|c_{A}|^{2}$ & $|c_{E_{1a}}|^{2}$ & $|c_{E_{2a}}|^{2}$ & $|c_{E_{B}}|^{2}$ & $|c_{E_{2b}}|^{2}$ & $|c_{E_{1b}}|^{2}$ & $l_\text{eff}(E_{1})$ & $l_\text{eff}(E_{2})$ 
\\ \hline
1 & -1.57 & 0.0  & 0.00 & 0.00 & 0.00 & 0.00 & 1.00 & 0.00 & +0.00 & -0.99 \\ \hline
2 & -1.35 & 0.0  & 0.00 & 0.00 & 1.00 & 0.00 & 0.00 & 0.00 & -0.00 & +0.99 \\ \hline
3 & -1.14 & -0.1 & 0.00 & 0.02 & 0.00 & 0.97 & 0.00 & 0.01 & +0.00 & +0.00 \\ \hline
4 & +0.45 & -2.96 & 0.00 & 0.97 & 0.00 & 0.01 & 0.00 & 0.02 & +0.95 & +0.00 \\ \hline
5 & +0.67 & -2.94 & 0.00 & 0.01 & 0.00 & 0.02 & 0.00 & 0.97 & -0.95 & +0.00 \\ \hline
6 & +2.96 & 0.0  & 1.00 & 0.00 & 0.00 & 0.00 & 0.00 & 0.00 & -0.00 & +0.00 \\ \hline
\end{tabular}
\caption{Results for fidelity check and effective angular momentum analysis for $C_{6}$. Having used the same parameters as in Fig. \ref{fig: linewidth} with $c_{ir}=0$. Band indices are assigned in ascending order of the real parts.}
\label{tab:c6_group}
\end{table*}

\subsubsection{Derivation of the effective optical angular momentum}
The IR basis is rotated by a geometric angle $\phi$ relative to this intrinsic basis, where the Hamiltonian is diagonal. The relation between the 2D block IR bases $\{|E_\pm\rangle$ and its corresponding intrinsic bases $|h_\pm\rangle$ can be expressed as
\begin{equation}
    \begin{pmatrix} |E_+\rangle \\ |E_-\rangle \end{pmatrix} = 
    \begin{pmatrix} \cos\phi & -\sin\phi \\ \sin\phi & \cos\phi \end{pmatrix}
    \begin{pmatrix} |h_+\rangle \\ |h_-\rangle \end{pmatrix}.
\end{equation}
Thus, $\phi$ can be calculated once the point group is fixed. The eigenstates will be further rotated by a dynamic angle $\theta$ if the 2D block $H_{E}$ has off-diagonal couplings. The dynamic angle is structure dependent and is specifically given by
\begin{equation}
    \theta = \arctan \left(\frac{|H_{E,12}|}{H_{E,11} - H_{E,22}}\right).
\end{equation}
Therefore, the total rotation of the eigenstates $|\psi_{\pm}\rangle$ relative to the IR basis, represented by $\Theta$, is given by
\begin{equation}
    \Theta = \theta + \phi.
\end{equation}
The observable $l_{\text{eff}}$ can thus be expressed as
\begin{equation}
    l_{\text{eff}} =  \cos(2 \Theta).
\end{equation}
Specific examples showing the results for $C_{6v}$ and $C_6$ are given in Sec. \ref{sec: quasibic}.
\subsection{Inclusion of Long-range Couplings}
\label{sec: appenberry}
Consider the two-dimensional triangular lattice in the Hermitian regime, with each lattice point in real space denoted as $\mathbf{R}_i$ and $i = 1, \dots, N$. Define the Hamiltonian as an isotropic decay form with arbitrarily long-range characteristics:
\begin{equation}
H = \sum_{i \neq j} t(|\mathbf{R}_i - \mathbf{R}_j|) c_i^\dagger c_j,
\end{equation}
where $t(|\mathbf{R}_i - \mathbf{R}_j|) \in \mathbb{R}, \quad t(|\mathbf{R}_i - \mathbf{R}_j|) = t(|\mathbf{R}_j - \mathbf{R}_i|)$, i.e., the coupling strength depends only on the distance $r_{ij} = |\mathbf{R}_i - \mathbf{R}_j|$, and the direction is isotropic.

Let $g \in C_{6v}$ denote any lattice rotation or mirror symmetry operation. We represent $g$ as a permutation matrix $D(g)$ in the lattice site basis. By applying $D(g)$, we have
\begin{equation}
\begin{split}
H' &= D(g) H D(g)^{-1} \\
   &= \sum_{i \neq j} t(|g \mathbf{R}_i - g \mathbf{R}_j|) c_{i}^\dagger c_{j} \\
   &= \sum_{i \neq j} t(|\mathbf{R}_i - \mathbf{R}_j|) c_i^\dagger c_j = H
\end{split}
\end{equation}
Here, we used the distance-preserving property
\begin{equation}
|g \mathbf{R}_i - g \mathbf{R}_j| = |\mathbf{R}_i - \mathbf{R}_j|, \quad \forall g \in C_{6v}.
\end{equation}

From above, we obtain $D(g) H D(g)^{-1} = H, \forall g \in C_{6v}$. This shows that the point group of the Hamiltonian, with all isotropic long-range couplings included, is also $C_{6v}$. That is to say, the symmetry of the long-range Hamiltonian is a natural extension of the nearest-neighbor lattice symmetry. The key to strictly preserving $C_6v$ symmetry is the isotropy of the coupling matrix. 

If $t(|\mathbf{R}_i - \mathbf{R}_j|)$ becomes anisotropic, $t(|\mathbf{R}_i - \mathbf{R}_j|, \theta_{ij})$, the rotation or mirror operations may no longer leave the Hamiltonian invariant, and the point group symmetry will be reduced. Nevertheless, the extension to lower-symmetry configurations, where band degeneracies and Berry-curvature hotspots may migrate to generic $k$-points, can be understood as a continuous evolution from the isotropic framework. 
\subsection{Implementation of Symmetry Perturbations in Experiments}
\label{sec: appenexp}
For experimental implementations, perturbations applied to symmetry modes are broadly classified into real and imaginary types, allowing for targeted manipulation through specific IR channels. Real perturbations primarily modify the real part of the energy spectrum to lift degeneracies and are typically achieved through geometric fine-tuning, anisotropic strain, environmental refractive index modulation, or the introduction of local scatterers. These methods enable precise control over properties like chiral or mirror symmetries and targeted mode mixing. Conversely, imaginary perturbations involve non-Hermitian gain and loss engineering, such as spatial pump patterning, to tune resonance thresholds and investigate exceptional points. Advanced control can also be realized via external fields to break TRS or through Floquet-type periodic modulations for synthesizing programmable dynamic couplings~\cite{haldane2008possible, wang2008reflection, jin2025towards, wang2025photoswitchable, yan2025topologically}. To implement these effectively, a standard experimental workflow begins with calibrating pure real perturbations to verify energy splitting, followed by applying pure imaginary perturbations to analyze mode selection and linewidths, and concludes with a combined complex perturbation sweep while continuously monitoring channel purity to systematically decouple frequency shifts from spectral broadening effects.

To illustrate these principles in a concrete physical system, we exemplify the direct correspondence between typical experimental actions target $C_{6v}$ symmetry channels in Table~\ref{tab:c6v_perturbations}. 
\begin{table*}[t]
\begin{tabular}{|l|l|l|l|l|l|}
\hline
Perturbation & Implementation & Representation & IR Channel & Primary Observables & $c_{\mathrm{ir}}$ \\\hline

\parbox[t]{2.6cm}{\raggedright Uniform Scaling} &
\parbox[t]{4.0cm}{\raggedright Uniform hole size adjustment, e.g.~\cite{chern2026dirac}} &
\parbox[t]{3.2cm}{\raggedright Diagonal energy shift} &
\parbox[t]{2.0cm}{\raggedright $A_2$ (Fully sym.)} &
\parbox[t]{4.0cm}{\raggedright $\mathrm{Re}(E)$ shift; degeneracies preserved} &
Real \\ \hline 

\parbox[t]{2.6cm}{\raggedright Alternating Geometric Distortion} &
\parbox[t]{4.0cm}{\raggedright Alternating long/short bonds or thickness between neighbors, e.g.~\cite{wu2015scheme}
} &
\parbox[t]{3.2cm}{\raggedright Diagonal differential term} &
\parbox[t]{2.0cm}{\raggedright $B_2$} &
\parbox[t]{4.0cm}{\raggedright Lifts specific degeneracies; opens bandgap} &
Real \\ \hline

\parbox[t]{2.6cm}{\raggedright In-Subspace Splitting ($E_1$)} &
\parbox[t]{4.0cm}{\raggedright Anisotropic deformation along one set of equivalent axes, e.g.~\cite{li2025observation}} &
\parbox[t]{3.2cm}{\raggedright $E_1$ subspace coupling ($2\times 2$ block)} &
\parbox[t]{2.0cm}{\raggedright $E_1$} &
\parbox[t]{4.0cm}{\raggedright $E_1$ doublet splits; alters polarization/angular profile} &
Real \\ \hline

\parbox[t]{2.6cm}{\raggedright In-Subspace Splitting ($E_2$)} &
\parbox[t]{4.0cm}{\raggedright Anisotropic deformation along another set of axes, e.g.~\cite{doiron2022realizing}} &
\parbox[t]{3.2cm}{\raggedright $E_2$ subspace coupling ($2\times 2$ block)} &
\parbox[t]{2.0cm}{\raggedright $E_2$} &
\parbox[t]{4.0cm}{\raggedright $E_2$ doublet splits} &
Real \\ \hline

\parbox[t]{2.6cm}{\raggedright Uniaxial Strain} &
\parbox[t]{4.0cm}{\raggedright Stretching or compression along $x$ or $y$ axis, e.g.~\cite{jamadi_direct_2020}
} &
\parbox[t]{3.2cm}{\raggedright Anisotropic real perturbation} &
\parbox[t]{2.0cm}{\raggedright $E_1 / E_2$ (orientation dep.)} &
\parbox[t]{4.0cm}{\raggedright Pronounced anisotropic splitting in $\mathrm{Re}(E)$} &
Real \\ \hline

\parbox[t]{2.6cm}{\raggedright Local Defect Scattering} &
\parbox[t]{4.0cm}{\raggedright Adding defect holes or side-coupled scatterers} &
\parbox[t]{3.2cm}{\raggedright Enhanced off-diagonal coupling} &
\parbox[t]{2.0cm}{\raggedright $E_1$ or $E_2$} &
\parbox[t]{4.0cm}{\raggedright Enhanced inter-mode mixing; pronounced avoided crossing} &
\parbox[t]{1.6cm}{\raggedright Real/Small Complex} \\ \hline

\parbox[t]{2.6cm}{\raggedright Absorption Loss Engineering} &
\parbox[t]{4.0cm}{\raggedright Patterned local metallic or absorptive layers, e.g.~\cite{salerno2022loss}} &
\parbox[t]{3.2cm}{\raggedright Anti-Hermitian (imaginary) term} &
\parbox[t]{2.0cm}{\raggedright $A_2 / B_2 / E_1 / E_2$} &
\parbox[t]{4.0cm}{\raggedright $\mathrm{Im}(E)$ (linewidth/$Q$) rearrangement; minor $\mathrm{Re}(E)$ shift} &
\parbox[t]{1.6cm}{\raggedright Pure~Imag. Complex} \\ \hline

\parbox[t]{2.6cm}{\raggedright Gain Engineering} &
\parbox[t]{4.0cm}{\raggedright Spatial optical pumping or gain doping, e.g.~\cite{kim2016direct}} &
\parbox[t]{3.2cm}{\raggedright Anti-Hermitian (imaginary) term} &
\parbox[t]{2.0cm}{\raggedright Injected per symmetry channel} &
\parbox[t]{4.0cm}{\raggedright Mode selection; staggered thresholding; $\mathrm{Im}(E)$ splitting} &
\parbox[t]{1.6cm}{\raggedright Pure~Imag. Complex} \\ \hline

\parbox[t]{2.6cm}{\raggedright Magneto-Optics / External Fields} &
\parbox[t]{4.0cm}{\raggedright Magnetic field + MO media; electro-optic modulation, e.g.~\cite{wang2009observation}} &
\parbox[t]{3.2cm}{\raggedright Diagonal/off-diagonal terms with phase} &
\parbox[t]{2.0cm}{\raggedright Direction-dependent projection} &
\parbox[t]{4.0cm}{\raggedright Asymmetric $\pm m$ shifts; non-reciprocal phenomena} &
\parbox[t]{1.6cm}{\raggedright Real, then Complex} \\\hline

\parbox[t]{2.6cm}{\raggedright Time-Modulation (Floquet)} &
\parbox[t]{4.0cm}{\raggedright Periodic modulation of index or coupling strength, e.g.~\cite{rechtsman_photonic_2013}} &
\parbox[t]{3.2cm}{\raggedright Effective complex phase engineering} &
\parbox[t]{2.0cm}{\raggedright Synthesized target channels} &
\parbox[t]{4.0cm}{\raggedright Simultaneous $\mathrm{Re}/\mathrm{Im}$ control; sideband generation} &
Complex \\
\hline
\end{tabular}
\caption{\label{tab:c6v_perturbations}Mapping of experimental perturbation schemes to $C_{6v}$ irreducible representation channels, model representations, and physical spectral observables.}
\end{table*}

\bibliography{bibliography}
\end{document}